\documentclass[10pt,conference,orivec]{llncs}

\usepackage{microtype}
\usepackage{wrapfig}
\usepackage{multirow}
\usepackage{amsmath}
\usepackage{amsfonts,graphicx,stmaryrd}
\usepackage{xspace}
\usepackage{cancel}
\usepackage{algorithm}
\usepackage[noend]{algpseudocode}
\usepackage{footnote}
\usepackage{comment}
\usepackage{balance}
\usepackage{color}
\usepackage{eucal}
\usepackage{wrapfig}
\usepackage{hyperref}
\usepackage{xcolor}
\newcommand{\deq}{\mathrel{\stackrel{\scriptscriptstyle\Delta}{=}}}
\algdef{SE}[DOWHILE]{Do}{doWhile}{\algorithmicdo}[1]{\algorithmicwhile\ #1}%

\usepackage[%
    font={small,sf},
    labelfont=bf,
    format=hang,    
    format=plain,
    margin=0pt,
    width=0.8\textwidth,
    singlelinecheck=on,
]{caption}
\usepackage[labelfont=bf,textfont=normalfont,singlelinecheck=on]{subcaption}

\usepackage{listings}

\definecolor{mGreen}{rgb}{0,0.6,0}
\definecolor{mGray}{rgb}{0.5,0.5,0.5}
\definecolor{mPurple}{rgb}{0.58,0,0.82}
\definecolor{backgroundColour}{rgb}{0.95,0.95,0.92}

\lstdefinelanguage{customC}%
  {morekeywords={auto,break,case,char,const,continue,default,do,double,%
      else,enum,extern,float,for,goto,if,int,long,register,return,%
      short,signed,sizeof,static,struct,switch,typedef,union,unsigned,uint,%
      void,volatile,while},%
  sensitive,%
  morecomment=[s]{/*}{*/},%
  morecomment=[l]//,
  morestring=[b]",%
  morestring=[m]',
  moredelim=*[directive]\#,%
  moredirectives={define,elif,else,endif,error,if,ifdef,ifndef,line,%
  include,pragma,undef,warning}%
}[keywords,comments,strings,directives]%

\lstdefinestyle{customc}{
  belowcaptionskip=1\baselineskip,
  xleftmargin=\parindent,
  language=customC,
  showstringspaces=false,
  basicstyle=\scriptsize\ttfamily,
  keywordstyle=\bfseries\color{green!40!black},
  commentstyle=\itshape\color{purple!40!black},
  identifierstyle=\color{black},
  stringstyle=\color{orange},
}

\newcommand{\tup}[1]
           {
             \relax\ifmmode
             \langle #1 \rangle
             \else $\langle$ #1 $\rangle$ \fi
           }

\usepackage{pgf}
\usepackage{tikz}
\usetikzlibrary{arrows,shapes,automata,petri,positioning,calc}
\usepackage{multicol}
\usepackage{amsmath,amsfonts}
\usepackage{graphicx}
\usepackage{textcomp}
\usepackage{xcolor}
\def\BibTeX{{\rm B\kern-.05em{\sc i\kern-.025em b}\kern-.08em
    T\kern-.1667em\lower.7ex\hbox{E}\kern-.125emX}}

\title{Direct Construction of Program Alignment Automata for Equivalence Checking}

\author{Manish Goyal\inst{1}\thanks{Both authors contributed equally to this research.}, Muqsit Azeem\inst{2}$^\star$, Kumar Madhukar\inst{3} and R. Venkatesh\inst{3}}
\institute{
	University of North Carolina at Chapel Hill, US \email{manishg@cs.unc.edu}
        \and
        Technical University of Munich, Germany \email{azeem@in.tum.de}
        \and
	TCS Research, Pune, India \email{\{kumar.madhukar, r.venky\}@tcs.com}
}

\begin{document}
\maketitle

\begin{abstract}
	The problem of checking whether two programs are semantically
	equivalent or not has a diverse range of applications, and is
	consequently of substantial importance. There are several techniques
	that address this problem, chiefly by constructing a product program
	that makes it easier to derive useful invariants. A novel addition to
	these is a technique that uses alignment predicates to align traces of
	the two programs, in order to construct a program alignment automaton.
	Being guided by predicates is not just beneficial in dealing with
	syntactic dissimilarities, but also in staying relevant to the
	property. However, there are also drawbacks of a trace-based technique.
	Obtaining traces that cover all program behaviors is difficult, and any
	under-approximation may lead to an incomplete product program.
	Moreover, an indirect construction of this kind is unaware of the
	missing behaviors, and has no control over the aforesaid
	incompleteness. This paper, addressing these concerns, presents an
	algorithm to construct the program alignment automaton directly instead
	of relying on traces.
\end{abstract}

\section{Introduction}
\label{sec:intro}

\begin{figure}
\centering
	\begin{minipage}{0.48\linewidth}
	\subcaptionbox{Function f\label{subfig:code-f}}[%
    0.7\linewidth\vspace
	{0.0cm} 
]%
{%
\begin{lstlisting}[label={lst:bitsflip-f-code}, caption={}, style=customc]
    void f(int* array, uint len)
    {
      for(i=0; i<len; i++)
        array[i] ^= 0xffffffff;
    }
\end{lstlisting}%
}%
	\end{minipage}
	\begin{minipage}{0.50\linewidth}
	\subcaptionbox{CFG for function f\label{subfig:cfg-f}}{\includegraphics[width=0.8\linewidth]{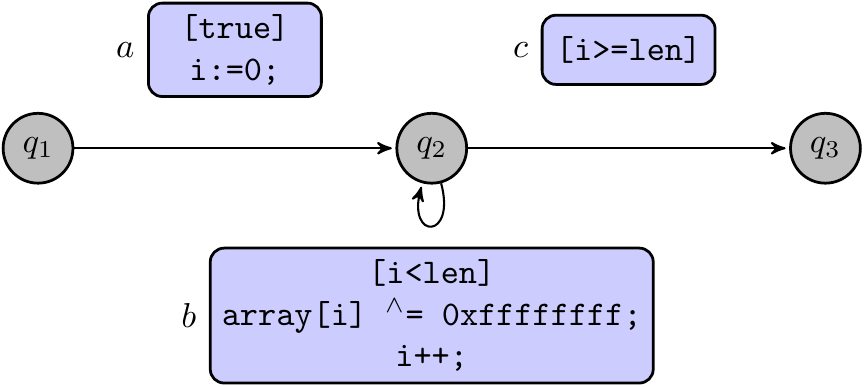}}
	\end{minipage}

	\vspace{0.2in}

	\begin{minipage}{0.48\linewidth}
	\subcaptionbox{Function g\label{subfig:code-g}}[%
    0.7\linewidth\vspace{0.3cm} 
]%
{%
\begin{lstlisting}[label={lst:bitsflip-g-code}, caption={}, style=customc]
    void g(int* array, uint len)
    {
      if(len%2==1) {
        *array ^= 0xffffffff;
        array++;
        len--;		
      }
      while(len) {
        *((long*)array) ^= 
	    0xffffffffffffffff;
        array += 2;
        len -= 2;
      }
    }
\end{lstlisting}%
}%
	\end{minipage}
	\begin{minipage}{0.50\linewidth}
	\subcaptionbox{CFG for function g\label{subfig:cfg-g}}{\includegraphics[width=0.8\linewidth]{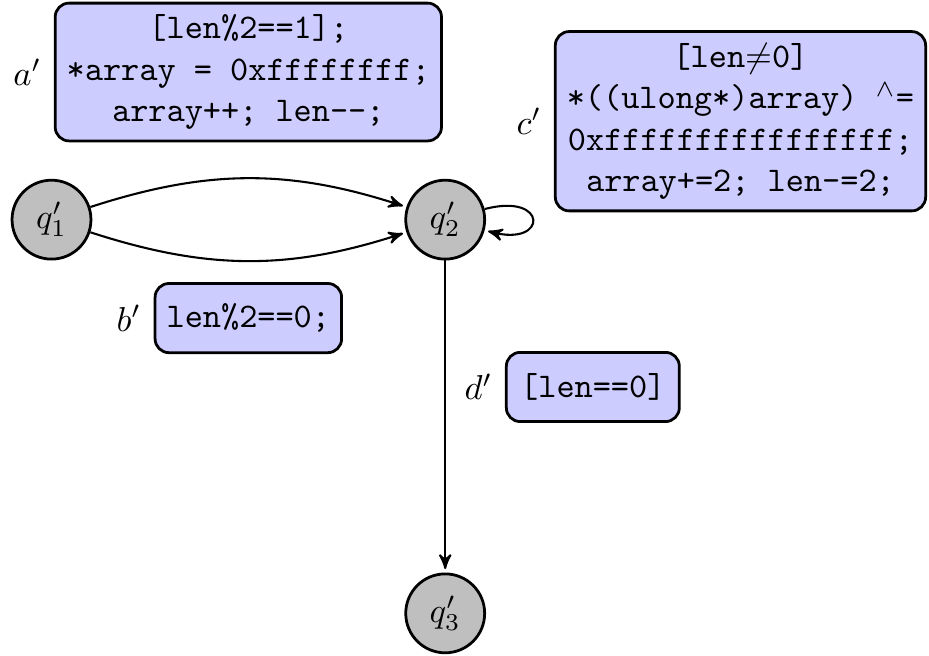}}
	\end{minipage}
\caption[]{Functions and their CFGs}
\label{fig:code-cfg-f-g}
\end{figure}

Checking equivalence of programs is an important problem due to its many diverse applications, including translation validation and compiler
correctness~\cite{10.1007/BFb0054170,GOLDBERG200553,10.1145/349299.349314},
code refactoring~\cite{10.1007/978-3-642-22110-1_55}, program
synthesis~\cite{10.1145/1168857.1168906}, hypersafety
verification~\cite{nier20hyper,ahv-cav19,pdsc-cav19},
superoptimization~\cite{10.1145/2451116.2451150,10.1145/3037697.3037754}, and software engineering education~\cite{Li:2016:MCB:2889160.2889204}, amongst many others.  In general,
depending on the application, the criteria for equivalence may be weaker or
stronger. For instance, the condition may be that all the observables including
the machine state (stack, heap, and registers) are equal, or that only a subset
of them are. Informally and broadly speaking, techniques that handle this
problem try to put the two programs together in a way that makes it easier to
justify the semantic equivalence. Note that one may always combine the programs
naively, like in a sequential composition where they are run one after the
other, but then arguing becomes difficult because it necessitates that every
component be analyzed fully. Consider an example (borrowed
from~\cite{spa-pldi19}) shown in Fig.~\ref{fig:code-cfg-f-g}. There are two
functions $f$ and $g$, both of which take two parameters as input:
\textit{array}, which points to an array of 32-bit integers, and \textit{len}
which stores the length of the array.  The function $f$ flips the bits of the
array elements by iterating over each array element, and function $g$ flips 64
bits from wherever the array is pointing to, and then moves the \textit{array}
pointer to the end of the flipped bits. In the beginning, however, $g$ checks
whether \textit{len} is odd and if so, flips only 32 bits for the first time,
and then continues flipping 64 at a time as described before. To establish that
the programs are semantically equivalent, one may simply put the two programs
together, one after another as sequential components of a single program, and
assert the equivalence condition at the end.  But, to analyze this combined
program, one must learn completely what is happening in $f$, and also in $g$,
and thereby conclude that they are indeed doing the same \emph{thing}.

The equivalence checking technique presented in~\cite{spa-pldi19}, on which we
build, takes two programs and set of test cases, and constructs a trace
alignment for every test case. The trace alignment is essentially a pairing of
states in the execution traces of the programs, corresponding to a test case.
This construction is guided by an alignment predicate that helps in pairing the
states semantically. The technique then builds the product program as a
\emph{program alignment automaton} (PAA), and then learns invariants for all
its states to establish the equivalence. In fact, the test cases are split into
two sets -- to be used for \emph{training} and \emph{testing} -- in the
beginning, and along with a set of candidate alignment predicates, a trace
alignment and a PAA are learned from the training data. In this setting, it
becomes important to ensure that the PAA does not overfit the training data.
Therefore, its viability is checked using the testing set. A PAA is acceptable
only if it soundly overapproximates the two programs, and is rejected
otherwise. In the latter case, the search for an acceptable PAA continues with
a different alignment predicate. Their technique benefits from choosing a good
alignment predicate that allows to capture all possible pairs of program
executions, including those from the testing set, even though it was learned
from the training data alone. 

The advantage of a \emph{semantic} alignment is that it can see through the
syntactic differences. However, there are also drawbacks of a trace-based
technique: \emph{a}) obtaining traces that cover all program behaviors is
difficult, and any under-approximation may lead to an incomplete product
program, and \emph{b}) an indirect construction of this kind is unaware of the
missing behaviors, and has no control over the aforesaid incompleteness.
Alternatively, there are techniques that do not need traces to arrive at a
product program, but they make assumptions that are strongly
limiting~\cite{DBLP:conf/aplas/DahiyaB17}. In this work, we propose an
algorithm for direct construction of PAA's, that has the goodness of being
guided by an alignment predicate, while still not needing any test cases or
unrealistic assumptions.


The core contribution of this paper is an algorithm for predicate guided
semantic program alignment without using traces, which we present in
Sect.~\ref{sec:algo}. This is followed by a step-by-step illustration of it on
the example of Fig.~\ref{fig:code-cfg-f-g}, in Sect.~\ref{sec:example}. We
present another illustrative run on an example involving arrays, which
emphasizes the usefulness of our direct construction over a trace-based
technique, in Sect.~\ref{sec:arrayinsert}. This is followed by a short note on
disjunctive invariants (in Sect.~\ref{sec:features}), a discussion of the
related work (in Sect.~\ref{sec:related}), and our concluding remarks (in
Sect.~\ref{sec:conc}).

\section{Equivalence Checking Algorithm}
\label{sec:algo}
\begin{wrapfigure}[8]{l}{0.47\textwidth}
  \begin{minipage}{0.47\textwidth}
	  \vspace{-0.6in}
	  \begin{algorithm}[H]
\caption{Equivalence checking}
\label{alg:equ-checking}
\begin{algorithmic}[1]
	\State $\mathcal{A}_{paa} \gets \mathit{paaConstruct}$ ($f, g, \mathcal{P}_{align}$){\label{alg1:paaconstruct}}
	\State $\mathcal{A}^{inv}_{paa} \gets \mathit{learnInvariants}$($\mathcal{A}_{paa}$) {\label{alg1:learninvariants}}
	\If{final-state invariant of $\mathcal{A}^{inv}_{paa} \Rightarrow$ equiv. prop{\label{alg1:checkproofob}}}
\State \textbf{return} \emph{equivalent}
\EndIf
\State \textbf{return} \emph{unknown}{\label{alg1:end}}
\end{algorithmic}
\end{algorithm}
\end{minipage}
\end{wrapfigure}

Alg.~\ref{alg:equ-checking} shows the procedure for checking equivalence
of two programs, $f$ and $g$. Given the programs and an alignment predicate
$\mathcal{P}_{align}$, it builds a program alignment automaton, learns
invariants for every state in the PAA, and then checks if the invariants in the
final state discharge the equivalence goal. The learned invariants need to be
consistent with the PAA, in the sense that if one picks an edge in the PAA,
then the invariants at the target state must follow from the ones at the
source, and the label on the chosen edge.

The inputs to the procedure $\mathit{paaConstruct}$ in
Alg.~\ref{alg:cons-paa} are automata $\mathcal{A}_{P_1}$ and
$\mathcal{A}_{P_2}$, which are CFGs of $f$ and $g$ resp., and an alignment
predicate $\mathcal{P}_{\mathit{align}}$. We assume that each program/function
has unique entry and exit state, $q_{init}$ and $q_{exit}$, akin to initial and
final state in an automaton. The procedure collects the states of both the
automata, and defines the states, $\mathcal{S}$, of the PAA to be their
product, i.e. each state in $\mathcal{S}$ is a tuple of two states $(q_i,
q_j)$, one from each automaton. The initial product state (which is simply the
product of the initial states) is marked reachable using the set
$\mathit{Reach}$, and the transitions ($\mathcal{T}$, an empty set in the
beginning) are populated one at a time, in a \emph{while} loop (lines
$7$-$20$).

In each iteration of the loop, a source state $(q_{i_1}, q_{{j_1}})$ is chosen
as any \emph{unvisited} state from the reachable set $Reach$ (as is marked
\emph{visited} immediately), along with a target state $(q_{i_2}, q_{{j_2}})$
from $\mathcal{S}$ (lines $7$-$9$). Then, the procedure derives a regular
expression denoting words in automaton $\mathcal{A}_{P_1}$, corresponding to
paths beginning at $q_{i_1}$ and ending at $q_{i_2}$. And, similarly, another
regular expressions for words in $\mathcal{A}_{P_2}$, for paths beginning at
$q_{j_1}$ and ending at $q_{j_2}$\footnote{These regular expressions between
program states need to be computed only once for every state combination, and can
be stored in a look-up table to avoid recomputation.}. At this
point, it discards this source-target pair if the regular expression
corresponds to an empty set in any of the automata. It also makes a discard if
the source and target states are the same, and the regular expression for any
of them is the empty word $\epsilon$. Intuitively, a discard of the former kind
means that the target is simply not reachable from the source in the product
program, whereas one of the latter kind denotes one of the programs is stuck in
a no-progress cycle. One may also discard aggressively, e.g. if the program
states in the target are not immediate neighbours of those in the source, but
this may come at the cost of completeness (feasible program behaviours missing
from the PAA).

The regular expressions are split over the top-level \emph{or} (+) to deal with the
different paths one at a time. This results into the sets $R_i$ and $R_j$,
obtained by splitting $rex_i$ and $rex_j$ resp., as shown in line $14$.
For every combination of paths (or, in other words, for every pair of regular
expression $r_i \in R_i$ and $r_j \in R_j$), the expressions are instantiated
by replacing $*$'s with symbolic constants $k_i$'s. The decision whether there
is a solution for the $k_i$'s in the instantiated expressions, such that an
appropriate edge labeling can be obtained, is left to an SMT solver (see
Sect.~\ref{subsec:concretize}). An edge is added between a source
and a target state only if the alignment predicate can be propagated along the
edge. Line $17$ of the pseudocode encodes this check. If an edge is added, the
target state is added to the $\mathit{Reach}$ set with an \emph{unvisited}
mark.

\begin{algorithm}[ht]
\caption{The program alignment automaton construction algorithm}
\label{alg:cons-paa}
\begin{algorithmic}[1]
	\Procedure{$\mathit{paaConstruct}$~}{$\mathcal{A}_{P_1}, \mathcal{A}_{P_2}, \mathcal{P}_{\mathit{align}}$}
	\State $S_1 \gets$ states of $\mathcal{A}_{P_1}$
	\State $S_2 \gets$ states of $\mathcal{A}_{P_2}$
\item[]
	\State $\mathcal{S} \gets$ \{($q_i,q_j$)\} where $q_i \in S_1, q_j \in S_2$ \Comment{\textcolor{gray}{set of product states}}
	\State $\mathit{Reach} \gets \{(q_{init_1},q_{init_2})\}$ \Comment{\textcolor{gray}{$q_{init_{i}}$ is the start state of automaton $\mathcal{A}_{P_i}$}}
	\State $\mathcal{T} \gets \emptyset$
\item[]
	\While{$\mathit{Reach}$ has a state $(q_{i_1}, q_{{j_1}})$, not yet marked \emph{visited}}
	\State mark $(q_{i_1}, q_{{j_1}})$ as \emph{visited}
	\For{$(q_{i_2}, q_{{j_2}}) \in \mathcal{S}$} \Comment{\textcolor{gray}{picking a target state to find transitions}}
	\State $rex_i \gets$ $\mathcal{L}$($\mathcal{A}_{P_1}$, with $q_{i_1}$ as initial and $q_{i_2}$ as final states)
	\State $rex_j \gets$ $\mathcal{L}$($\mathcal{A}_{P_2}$, with $q_{j_1}$ as initial and $q_{j_2}$ as final states)
	\Statex \Comment{\textcolor{gray}{discard the state-pair in case of no paths, or if there is a no-progress cycle}}
	\State $\mathbf{next}$ if ($rex_i$ = $\emptyset$) or ($rex_j$ = $\emptyset$)
	\State $\mathbf{next}$ if ($q_{i_1}$ = $q_{i_2}$ and $q_{j_1}$ = $q_{j_2}$) and ($rex_i = \{\epsilon\}$ or $rex_j = \{\epsilon\}$)
	\item[]
	\State $R_i \gets \mathit{split}$($rex_i$); $R_j \gets \mathit{split}$($rex_j$) 
	\For{$(r_i,r_j) \in R_i \times R_j$}
	\State ${r^c}_i, {r^c}_j$ $\gets \mathit{instantiate}$($r_i, r_j$)  \Comment{\textcolor{gray}{replace $\ast$ with constants $k_i$'s}}
	\State find min $k_i$'s: $\mathcal{P}_{\mathit{align}} \wedge {r^c}_i \wedge {r^c}_j \Rightarrow \overline{\mathcal{P}}_{\mathit{align}}$ \Comment{\textcolor{gray}{$\overline{~\cdot~}$ denotes next-state}}
	\If {a solution is found} 
	\State $\mathcal{T} \gets \mathcal{T} ~~ \cup ~~ (q_{i_1}, q_{j_1}) \xrightarrow{{r^c}_i;{r^c}_j} (q_{i_2}, q_{j_2})$
	\State $Reach \gets Reach \cup \{(q_{i_2}, q_{j_2})\}$
	\EndIf
	\EndFor
	\EndFor
	\EndWhile
\item[]
	\State $\mathcal{S} \gets Reach$ \Comment{\textcolor{gray}{unreached states are removed from $\mathcal{S}$ in the end}}
	\State $\mathit{simplify}(\mathcal{T}, \mathcal{S})$
\EndProcedure
\end{algorithmic}
\end{algorithm}

The \emph{while} loop exits when all the reachable states have been marked
\emph{visited}.  At this point the unreached states are removed from $S$, and
the resulting set along with the set of transitions $\mathcal{T}$, describes
the program alignment automaton obtained thus. The resulting PAA is also
simplified, in a manner similar to~\cite{spa-pldi19}, as explained in
Sect.~\ref{subsec:reduce}.

The usefulness of an alignment predicate reflects in how well it helps align
the programs and discharge the equivalence property. For example, if an
alignment predicate only helps to align the initial and the final states, and
no other state in between, it does not make the proof any more easier than
completely analysing the programs independently. Finding good alignment
predicates is thus important, but also quite challenging at the same
time~\cite{spa-pldi19}.  Though we do not address this problem here, we believe
that data- and syntax-guided techniques can be quite helpful in making this
practicable. For example, the technique in~\cite{spa-pldi19} learns a set of
candidate alignment predicates from the training data.  Similarly, one may
construct a grammar and sample these candidates automatically from the program
source following the ideas
of~\cite{freqhorn-fmcad17,freqhorn-sas18,freqhorn-cav19}.

\subsection{Propagating preconditions along transitions}
\label{rem:input-pred}

	In addition to the alignment predicate, there are also predicates that
	capture the \emph{preconditions} under which we are checking
	equivalence.  This could, for instance, be a predicate equating the
	input variables of the two programs. Let $\mathcal{P}_{\mathit{input}}$
	denote a set of such predicates. When a transition is added to the PAA,
	the predicates in this set are also propagated to the target state if
	they hold there. These predicates help in the propagating the alignment
	predicate by strengthening the premise of the check in line $17$. If
	the alignment predicate can be propagated along an edge without the
	help of these input predicates, then the edge is added as it is.
	Otherwise, if it is propagated with the assistance of the input
	predicates, then the edge is marked (as ``dependent on an input
	predicate'') before it is added to a set of marked transitions,
	$\mathcal{T}_m$ (instead of $\mathcal{T}$). Once the PAA construction
	is over, if an input predicate has not been propagated to any state, we
	remove all marked edges from the state that are dependent on that
	predicate.

	The reason we separate the marked transitions from the unmarked ones is
	to avoid backtracking.  A predicate $p \in
	\mathcal{P}_{\mathit{input}}$ holds at a non-initial state $s$ only if
	it is preserved along all paths that reach $s$. At an intermediate
	stage in the construction, even if $p$ holds at $s$, it may later be
	discovered to not hold there.  However, if $p$ was used at that stage
	to propagate the alignment predicate, we would need to remove that edge
	and backtrack. Marking such transitions and keeping them separately
	allows us to get rid of all of them, at once, in the end.

\subsection{Reduction of Program Alignment Automaton}
\label{subsec:reduce}

The procedure $\mathit{simplify~}(\mathcal{T},\mathcal{S})$ reduces the program alignment
automaton for $\mathcal{P}_{align}$ by repeatedly applying the following two
reductions, as long as they have some effect.

\begin{enumerate}
    \item $\mathit{RemoveStates~}$ removes every state $s$, other than the
	    initial and the final state, that does not have a self-loop.
		Essentially, it replaces each pair of transitions $s_k
		\xrightarrow{P;Q} s$ and $s \xrightarrow{P';Q'} s_l$, where $s$
		does not have a self-loop, with a transition $s_k
		\xrightarrow{} s_l$ labeled with $PP';QQ'$.
    \item $\mathit{RemoveTransitions~}$ removes transitions of the form $s
	    \xrightarrow{P';Q'} s_k$, if there is a transition $s
		\xrightarrow{P;Q} s_l$ where $P$ is a prefix of $P'$ and $Q$
		is a prefix of $Q'$.
\end{enumerate}

\subsection{Concretization of regular expressions}
\label{subsec:concretize}

While adding transitions in the PAA, we employ a solver to compute valid
solutions of $k_i$'s in the instantiated regular expressions. In this
subsection, we describe why it is sufficient to find these instantiations such
that they account for all program behaviours. In our PAA construction, there
can be three types of the transition labels $P;Q$.

\begin{enumerate}
    \item Both $P$ and $Q$ contain loop blocks, i.e. the label is of form
	    $r_i r_i^{k_1}; r_jr_j^{k_2}$ where $r_i$ and $r_j$ are the blocks
		denoting loops in respective functions. In this case, we find
		the minimum values of $k_1$ and $k_2$ such that $k_1+1$
		iterations of $r_i$ are aligned with $k_2+1$ iterations of
		$r_j$. By not considering their minimum values, we will be
		unable to account for the behaviours with smaller number of
		loop iterations. However, minimum values automatically
		accommodate behaviors with higher number of loop iterations.
    
    \item Only one of $P$ and $Q$ has a loop block, i.e. the label is
	    of one of the forms $r_{i-1}r_i^{k};r_j$, $r_i^{k}r_{i+1};r_j$,
		$r_i;r_j^{k}r_{j+1}$ or $r_i;r_{j-1}r_j^{k}$, where the
		expression with superscript $k$ denotes the loop block. In this
		case, we check if there exists a value of $k$ such that the transition preserves the validity of alignment predicate.
		Intuitively, this value of $k$ determines how many iterations
		of loop in one function are aligned with a non-loop block in
		other function.
    
    \item Neither $P$ nor $Q$ has a loop block, i.e. the label is
	    $r_i;r_j$. Here, merely checking that taking this
		transition does not violate the alignment predicate is
		sufficient.
\end{enumerate}

\section{Illustrative run on an example}
\label{sec:example}

We use the example in Fig.~\ref{fig:code-cfg-f-g} to illustrate
Alg.~\ref{alg:cons-paa}.  The inputs to $\mathit{paaConstruct}$ are the
automata shown in figures~\ref{subfig:cfg-f} and~\ref{subfig:cfg-g}, and an
alignment predicate $\mathcal{P}_{align}$: $array+4i=array'$. The sets $S_1$
and $S_2$ are $\{q_1,~q_2,~q_3\}$ and $\{q'_1,~q'_2,~q'_3\}$ (resp.), and thus
the PAA has nine possible states
$\{q_1q'_1,~q_1q'_2,~\ldots,~q_3q'_2,~q_3q'_3\}$. We often denote the product
state $(q_i, q_j)$ as $q_iq_j$. The states $q_1q'_1$ and $q_3q'_3$ are marked
as initial and final, resp. We assume that the alignment predicate holds in the
initial state, without evaluating whether it actually holds or not.  As
described in Sect~\ref{rem:input-pred}, we also have a set of input predicates
(omitted from Alg.~\ref{alg:cons-paa} for ease of exposition) that hold in the
beginning.  In this example, it is the set $\{array = array', len = len',
\omega = \omega'\}$, where $\omega$ denotes the heap state. The predicate
$\omega = \omega'$ is the precondition that the programs execute from the same
heap state. Input predicates holds at the initial state, and at any subsequent
state unless a transition flips their truth value.

\vspace{-0.35in}
\begin{table}
\caption{Computing PAA transitions}
\label{fig:transition_table}
\scriptsize
    \centering
\begin{tabular} { | p {0.7 cm} | p {2.2 cm} | p {2.5 cm} | p {4.2 cm} | p {1.0 cm} | p {0.8 cm} | }
\hline
Row & ~Transition & ~~~~~~~~~Regex & ~~~~~~~Instantiation & Label & Set\\
\hline
	1 & $q_1q'_1 \xrightarrow{} q_2q'_1$ & $ab^*;\epsilon$ & $ab^{k_1};\epsilon$\hfill${\{{\scriptstyle k_1=0}\}}$ & $a;\epsilon$ & $\mathcal{T}_m$ \\
	2 & $q_1q'_1 \xrightarrow{} q_1q'_2$ & $\epsilon;b'c'^*+a'c'^*$ & $\epsilon;b'c'^{k_1}$\hfill$\{{\scriptstyle k_1=0}\}$ & $\epsilon;b'$ & $\mathcal{T}$\\
  &                                  &                                    & $\epsilon;a'c'^{k_2}$\hfill$\{{\scriptstyle k_2=-}\}$ & $-$ & $-$\\
	3 & $q_1q'_2 \xrightarrow{} q_2q'_1$ & $ab^*;\emptyset$ & no path from $q'_2$ to $q'_1$ & $-$ & $-$\\
	4 & $q_1q'_1 \xrightarrow{} q_2q'_2$ & $ab^*;b'c'^*+a'c'^* $ & $ab^{k_1};b'c'^{k_2}$\hfill$\{{\scriptstyle k_1=0, k_2=0}\}$ & $a;b'$ & $\mathcal{T}_m$ \\
	&                                    &                     & $ab^{k_3};a'c'^{k_4}$\hfill$\{{\scriptstyle k_3=1, k_4=0}\}$ & $ab;a'$ & $\mathcal{T}_m$\\
	5 & $q_2q'_1 \xrightarrow{} q_2q'_1$ & $b^*;\epsilon$ & $\text{no-progress cycle}$\hfill${\{{\scriptstyle -}\}}$ & $-$ & $-$\\
	6 & $q_2q'_1 \xrightarrow{} q_2q'_2$ & $b^*;a'c'^*+b'c'^*$ &  $b^{k_1};a'c'^{k_2}$\hfill$\{{\scriptstyle k_1=1, k_2=0}\}$ & $b;a'$ & $\mathcal{T}$ \\
	&                                  &                                           & $b^{k_3};b'c'^{k_4}$\hfill$\{{\scriptstyle k_3=0, k_4=0}\}$ & $\epsilon;b'$  & $\mathcal{T}$ \\
	7 & $q_2q'_1 \xrightarrow{} q_3q'_2$ & $b^*c;a'c'^*+b'c'^*$ &  $b^{k_1}c;a'c'^{k_2}$\hfill$\{{\scriptstyle k_1=1, k_2=0}\}$ & $bc;a'$ & $\mathcal{T}$\\
	&                                  &                                           & $b^{k_3}c;b'c'^{k_4}$\hfill$\{{\scriptstyle k_3=0, k_4=0}\}$ & $c;b'$  & $\mathcal{T}$\\
	8 & $q_2q'_1 \xrightarrow{} q_3q'_1$ & $b^*c;\epsilon$ & $b^{k_1}c;\epsilon$\hfill${\{{\scriptstyle k_1=0}\}}$ & $c;\epsilon$ & $\mathcal{T}$ \\
	9 & $q_1q'_2 \xrightarrow{} q_1q'_3$ & $\epsilon;c'^*d'$ & $\epsilon;c'^{k_1}d'$\hfill${\{{\scriptstyle k_1=0}\}}$ & $\epsilon;d'$ & $\mathcal{T}$\\
	10 & $q_1q'_2 \xrightarrow{} q_2q'_2$ & $ab^*;c'^*$ & $ab^{k_1};c'^{k_2}$\hfill${\{{\scriptstyle k_1=0, k_2=0}\}}$ & $a;\epsilon$ & $\mathcal{T}_m$ \\
	11 & $q_1q'_2 \xrightarrow{} q_2q'_3$ & $ab^*;c'^*d'$ & $ab^{k_1};c'^{k_2}d'$\hfill${\{{\scriptstyle k_1=0, k_2=0}\}}$ & $a;d'$ & $\mathcal{T}_m$\\
    12 & $q_3q'_1 \xrightarrow{} q_3q'_2$ & $\epsilon;a'c'^*+b'c'^*$ &  $\epsilon;a'c'^{k_1}$\hfill$\{{\scriptstyle k_1=-}\}$ & $-$ & $-$\\
	&                                  &                                           & $\epsilon;b'c'^{k_2}$\hfill$\{{\scriptstyle k_2=0}\}$ & $\epsilon;b'$ & $\mathcal{T}$\\
	13 & $q_1q'_3 \xrightarrow{} q_2q'_3$ & $ab^*;\epsilon$ & $ab^{k_1};\epsilon$\hfill${\{{\scriptstyle k_1=0}\}}$ & $a;\epsilon$ & $\mathcal{T}_m$ \\
	14 & $q_2q'_2 \xrightarrow{} q_2q'_2$ & $b^*;c'^*$ & $bb^{k_1};c'c'^{k_2}$\hfill${\{{\scriptstyle k_1=1, k_2=0}\}}$ & $bb;c'$ & $\mathcal{T}$ \\
	15 & $q_2q'_2 \xrightarrow{} q_2q'_3$ & $b^*;c'^*d'$ & $b^{k_1};c'^{k_2}d'$\hfill${\{{\scriptstyle k_1=0, k_2=0}\}}$ & $\epsilon;d'$ & $\mathcal{T}$\\
	16 & $q_2q'_2 \xrightarrow{} q_3q'_2$ & $b^*c;c'^*$ & $b^{k_1}c;c'^{k_2}$\hfill${\{{\scriptstyle k_1=0, k_2=0}\}}$ & $c;\epsilon$ & $\mathcal{T}$\\
	17 & $q_2q'_2 \xrightarrow{} q_3q'_3$ & $b^*c;c'^*d'$ & $b^{k_1}c;c'^{k_2}d'$\hfill${\{{\scriptstyle k_1=0, k_2=0}\}}$ & $c;d'$ & $\mathcal{T}$\\
	18 & $q_2q'_3 \xrightarrow{} q_3q'_3$ & $b^*c;\epsilon$ & $b^{k_1}c;\epsilon$\hfill${\{{\scriptstyle k_1=0}\}}$ & $c;\epsilon$ & $\mathcal{T}$ \\
	19 & $q_3q'_2 \xrightarrow{} q_3q'_3$ & $\epsilon;c'^*d'$ & $\epsilon;c'^{k_1}d'$\hfill${\{{\scriptstyle k_1=0}\}}$ & $\epsilon;d'$ & $\mathcal{T}$\\
	20 & $q_1q'_3 \xrightarrow{} q_1q'_3$ & $\epsilon;\epsilon$ & $\text{no-progress cycle}$\hfill${\{{\scriptstyle -}\}}$ & $-$ & $-$\\
\hline
\end{tabular}
\end{table}

We mark the initial state $q_1q'_1$ as \emph{reachable} by initializing the set
$Reach$ with it (line $5$).  We also initialize the transition set
$\mathcal{T}$ to be an empty set. The process of adding a transition begins by
picking two states: a \emph{source} state from the $Reach$, and a \emph{target}
from $\mathcal{S}$.
Table~\ref{fig:transition_table} shows all the transitions that were added by
the algorithm. In what follows, we describe a few interesting cases in details.

\textbf{Single transition} Consider the pair of states $q_1q'_1 \in Reach$ and
$q_2q'_1 \in \mathcal{S}$ at the entry 1 in Table~\ref{fig:transition_table}.
We mark $q_1q'_1$ as \emph{visited} before proceeding (line $8$). The regular
expression $ab^*$ denotes all the words beginning at $q_1$ and ending at $q_2$
in Fig.~\ref{subfig:cfg-f} (line $10$). Similarly, $\epsilon$ denotes the
words starting at $q'_1$ and ending at $q'_1$ in Fig.~\ref{subfig:cfg-g}
(line $11$). Since these expressions do not have a top-level or (denoted by
`$+$'), we just get two singleton sets -- $\{ab^*\}$ and $\{\epsilon\}$ -- in
line $14$. Recall that, by assumption, both $\mathcal{P}_{align}$ and
$\mathcal{P}_{input}$ hold at $q_1q'_1$.  We employ an SMT solver to find an
instantiation, if one exists, of $ab^{k_1};\epsilon$, such that
$\mathcal{P}_{align}$ retains its truth value at $q_2q'_1$ after taking the
transition (lines $16-19$). In particular, we solve the query $array+4i=array'
\land ab^{k_1} \land \epsilon \implies
\overline{array}+4\overline{i}=\overline{array}'$ for the minimum value of
$k_1$. As the solver does not find any satisfying assignment, we try to solve
the query by adding $\mathcal{P}_{input}$ to the premise. The solver now
returns $0$ as the solution which results into a transition $q_1q'_1
\xrightarrow{a;\epsilon} q_2q'_1$. We add this transition to $\mathcal{T}_m$
(see Sect.~\ref{rem:input-pred}) and add $q_2q'_1$ to $Reach$ (lines $20-22$).
Further, since the basic block $a$ in Fig.~\ref{subfig:cfg-f} does not affect
the truth value of $\mathcal{P}_{input}$, we propagate $\mathcal{P}_{input}$ to
$q_2q'_1$ through this transition by making a similar query to the solver.

\begin{wrapfigure}[16]{l}{0.38\textwidth}
\centering
\includegraphics[width=\linewidth]{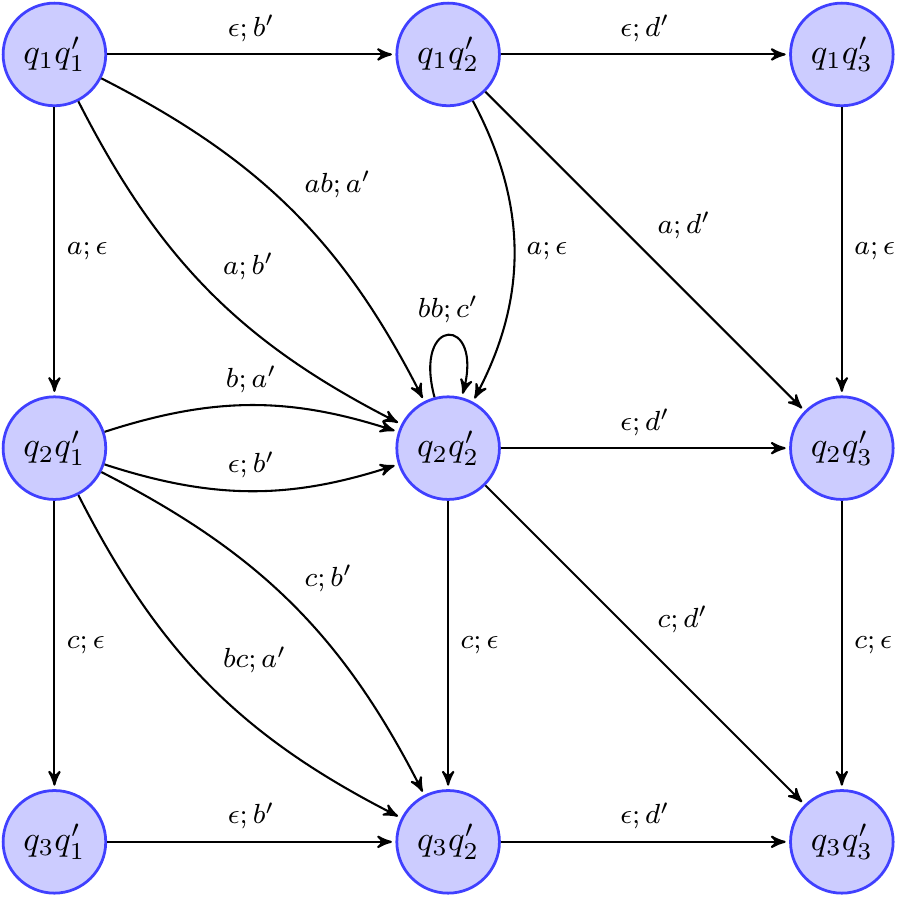}
\caption{PAA for Fig.~\ref{fig:code-cfg-f-g}}
    \label{fig:full-paa}
\end{wrapfigure}

Let us pick another pair of states: $q_1q'_1$ and $q_1q'_2$, the second entry in Table~\ref{fig:transition_table}. The regular expressions denoting the words
between component states are $\epsilon$ and $b'c'^* + a'c'^*$, and splitting
gives two sets - $\{\epsilon\}, \{b'c'^*; a'c'^*\}$. We solve two queries in
order to obtain their instantiations: (i) $array+4i=array' \land \epsilon \land
b'c'^{k_1} \implies \overline{array}+4\overline{i}=\overline{array}'$, and (ii)
$\mathcal{P}_{\mathit{input}}\land array+4i=array' \land \epsilon \land
a'c'^{k_2} \implies \overline{array}+4\overline{i}=\overline{array}'$. For the
first query, the solver provides $k_1=0$. We add a transition $q_1q'_1
\xrightarrow{\epsilon;b'} q_1q'_2$ to $\mathcal{T}$ and add $q_1q'_2$ to
$Reach$. Notice that we added $\mathcal{P}_{input}$ to the premise in second
query after we observed that the solver could not find an instantiation without
$\mathcal{P}_{input}$. The second query could not be solved, even with
$\mathcal{P}_{input}$.  We propagate $\mathcal{P}_{input}$ to $q_1q'_2$, as it
is not affected by the edge labeled $\epsilon;b'$.

\textbf{Discarding a pair of states} Consider a pair $q_1q'_2$ and $q_2q'_1$ at
entry $3$ in Table~\ref{fig:transition_table}. Note that there is a path from
$q_1$ to $q_2$ in the automaton in Fig.~\ref{subfig:cfg-f} but there is no
path from $q'_2$ to $q'_1$ in ~\ref{subfig:cfg-g}. So, we discard this pair
since no transition can be added from $q_1q'_2$ to $q_2q'_1$ (line $12$ in
Alg.~\ref{alg:cons-paa}). Consider another pair, $q_2q'_1$ and $q_2q'_1$, at
entry $5$. The associated regex $b^*;\epsilon$ represents all the words
starting at $q_2q'_1$ and ending at $q_2q'_1$. As this expression would result
into a \emph{no-progress cycle} (the states does not change in any of the
components, and at least one of the expressions is $\epsilon$) at $q_2q'_1$, we
discard this pair (line $13$ in Alg.~\ref{alg:cons-paa}).

In this example, we only look for the pairs where component states are
immediate neighbors in respective automaton. For instance, we do not look for a
transition between $q_1q'_1$ and $q_1q'_3$ because $q'_1$ and $q'_3$ are not
immediate neighbors in Fig.~\ref{subfig:cfg-g}. Such optimizations, in general,
may lead to loss of behaviours.

\textbf{Multiple transitions} 
For the pair $q_2q'_1$ and $q_2q'_2$ at entry $6$, the regular expressions are
$b^*$ and $a'c'^*+b'c'^*$. Splitting gives us the sets: $\{b^*;a'c'^*\}$ and
$\{b^*;b'c'^*\}$. The solver returns $k_1=1, k_2=0$ as the instantiation of
$b^{k_1};a'c'^{k_2}$, which gives the edge label $b;a'$. For the other
component, $b^{k_3};b'c'^{k_4}$, the transition label becomes $\epsilon;b'$ as
the solver return $k_3=0, k_4=0$. These were obtained with
$\mathcal{P}_{input}$ in the premise, and thus, are added to $\mathcal{T}$.
The state $q_2q'_2$ is added to $Reach$. Observe that the truth values of
$array=array'$ and $len=len'$ are affected by the blocks $b$ and $a'$,
therefore these are dropped at the \emph{target} state. However, since $\omega
= \omega'$ is still unaffected, we propagate it to $q_2q'_2$.

\textbf{Self-loop} The transition $b^*;c'^*$ at entry 14 corresponds to a
self-loop at the state $q_2q'_2$. We get $k_1=2,k_2=1$ as the instantiation of
$bb^{k_1};c'c'^{k_2}$ using the solver, and add a self-loop with label $bb;c'$
at $q_2q'_2$. Informally, it shows that a transition with two iterations of $b$
and one iteration of $c'$ preserves the satisfiability of $\mathcal{P}_{align}$
at $q_2q_{2}'$. Note that we do not enquire for minimum values of $k_i$'s in
the case of $b^{k_1};c'^{k_2}$, because the minimum ($k_1=0,k_2=0$) corresponds
to a no-progress cycle. The query is suitably modified for self-loops to ensure
progress. 

\begin{figure}
\centering
\begin{subfigure}[t]{0.45\textwidth}
\includegraphics[width=\linewidth]{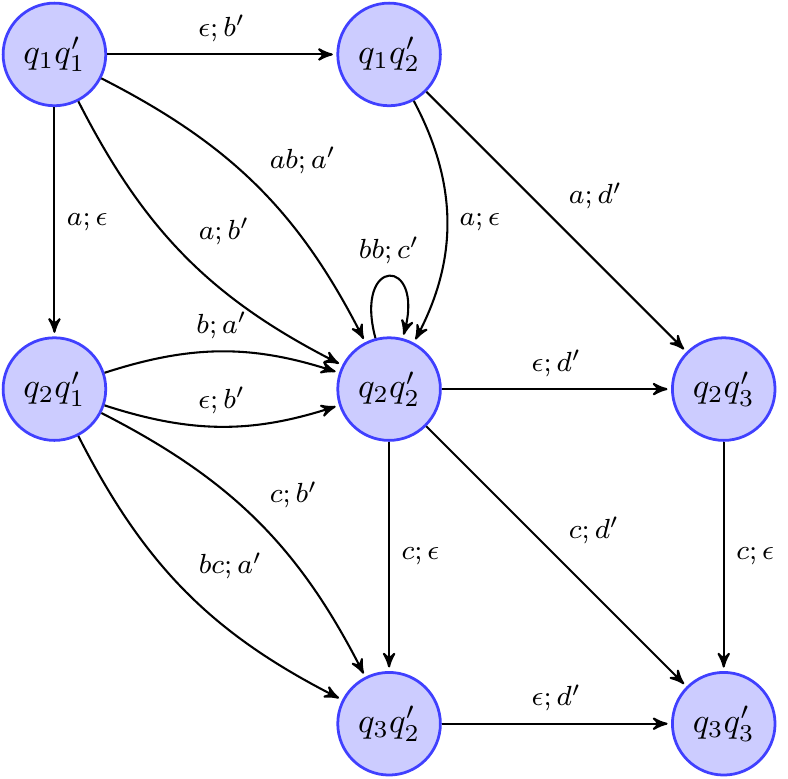}
 \caption{After removing states $q_1q'_3$ and $q_3q'_1$\label{subfig:simplify-1}}
        \end{subfigure}
 \quad
\begin{subfigure}[t]{0.45\textwidth}
\includegraphics[width=\linewidth]{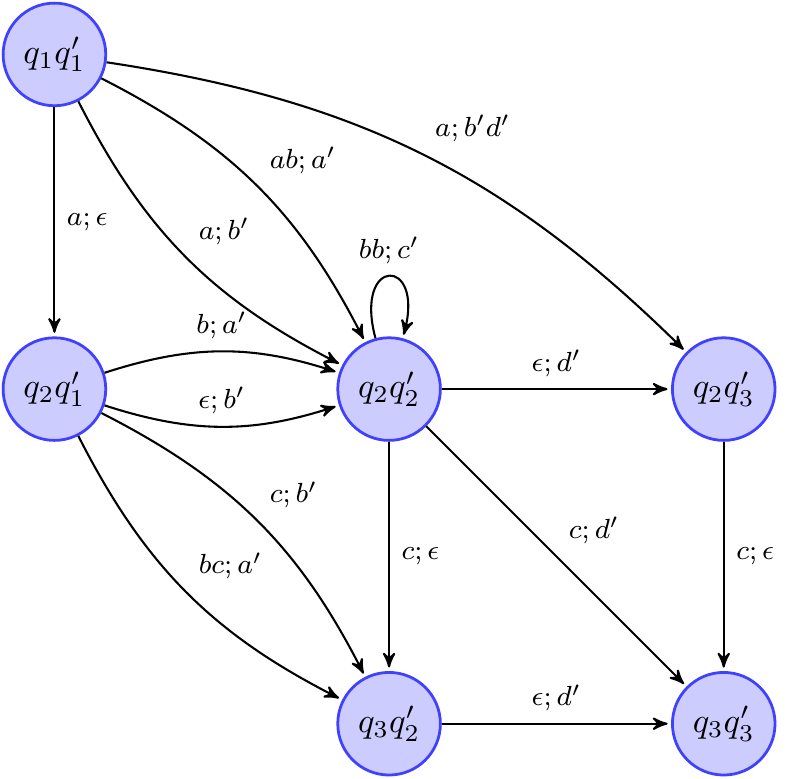}
 \caption{After removing state $q_1q'_2$\label{subfig:simplify-2}}
        \end{subfigure}

 \caption{Reduction of PAA in Figure~\ref{fig:full-paa}}
     \label{fig:simplification}
\end{figure}

\vspace{-0.3in}

\begin{figure}
    \centering
    
\begin{subfigure}[t]{0.36\textwidth}
\includegraphics[width=\linewidth]{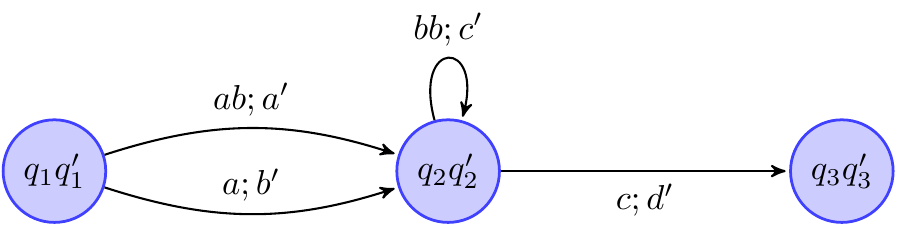}
 \caption{proposed construction\label{subfig:our-paa}}		
 	\end{subfigure}
 \quad
 \hspace{20pt}
\begin{subfigure}[t]{0.45\textwidth}
\includegraphics[width=\linewidth]{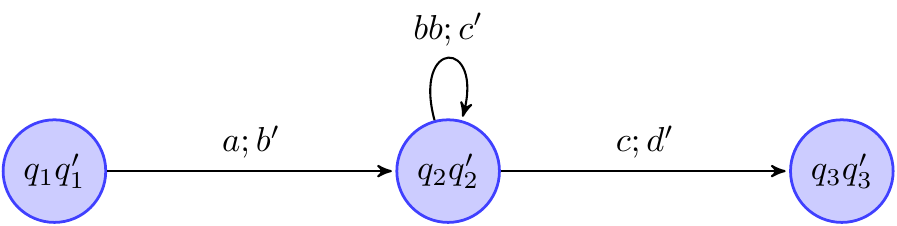}
	\caption{using~\cite{spa-pldi19}, and only even \texttt{len} traces\label{subfig:pldi-paa}}		
 	\end{subfigure}
\caption{Final program alignment automaton}
    \label{fig:final-paas}
\end{figure}

Once the while loop ends, the unreachable states are removed from
$\mathcal{S}$, and the valid transition of $\mathcal{T}_m$ are added to
$\mathcal{T}$. A marked transition is valid if the input predicates used in the
premise continue to be available at the source state in the end. The PAA thus
constructed, shown in Fig.~\ref{fig:code-cfg-f-g}, is then simplified. For
instance, we can remove state $q_1q'_3$ by replacing transitions $q_1q'_2
\xrightarrow{\epsilon;d'} q_1q'_3$ and $q_1q'_3 \xrightarrow{a;\epsilon}
q_2q'_3$ with a transition $q_1q'_2 \xrightarrow{a;d'} q_2q'_3$ which is
already present. The reduced PAA is shown in Fig.~\ref{subfig:simplify-1}. In
a similar manner, state $q_1q'_2$ is removed by replacing - (i) the transitions
$q_1q'_1 \xrightarrow{\epsilon;b'} q_1q'_2$ and $q_1q'_2 \xrightarrow{a;d'}
q_2q'_3$ with a transition $q_1q'_1 \xrightarrow{a;b'd'} q_2q'_3$, (ii) the
transitions $q_1q'_1 \xrightarrow{\epsilon;b'} q_1q'_2$ and $q_1q'_2
\xrightarrow{a;\epsilon} q_2q'_2$ with a transition $q_1q'_1 \xrightarrow{a;b'}
q_2q'_2$.  Next, we remove the transition
$q_1q'_1 \xrightarrow{a;b'd'} q_2q'_3$ because there exists a transition
$q_1q'_1 \xrightarrow{a;b'} q_2q'_2$ where $a$ is a prefix of $a$ and $b'$ is a
prefix of $b'd'$. We keep applying these reductions until the PAA can not be
simplified further. The final PAA is shown in Fig.~\ref{subfig:our-paa}. This
is exactly same as the PAA obtained by the technique in~\cite{spa-pldi19}.
However, since their technique depends on test cases, if the training set had
only even \texttt{len} cases (for example), they would have ended up with a
different PAA, shown in Fig.~\ref{subfig:pldi-paa}. Observe that this PAA
does not have a transition corresponding to edge $a'$\texttt{[len\%2 = 1]} in
Fig.~\ref{subfig:cfg-g}, and therefore does not overapproximate all possible
behaviors.

\subsection{Learning Invariants and Discharging Proof Obligations}
\label{subsec:learnandprove}
Though we do not have any contributions here, we illustrate how this is done
(in~\cite{spa-pldi19}) to make the paper self-contained.

Once a PAA is constructed, invariants are learned for each state. These
invariants must be consistent with PAA i.e, for each transition $s
\xrightarrow{P;Q} t$, if $\phi_s$ and $\phi_t$ are the invariants at state $s$
and $t$ respectively, then, $\{\phi_s\}~P;Q~\{\phi_t\}$ must be valid.  The aim
is to learn sufficiently strong invariants at the final state, so that the
equivalence property can be discharged. There are several techniques that have
been proposed to learn such invariants~\cite{spa-pldi19,DBLP:conf/aplas/DahiyaB17}, including those that aim to learn them from the
program's syntactic source, e.g.~\cite{freqhorn-fmcad17}.

It must be first argued that the constructed PAA overapproximates all program
behaviors. Consider the initial state $q_1q'_1$: the state $q_1$ in $f$ has one
outgoing transition with its guard predicate as $true$ (say, $\alpha$),
whereas, $q'_1$ has two outgoing transitions with guard predicates $len\%2=0$
($\beta$) and $len\%2=1$ ($\gamma$). Hence there are two possible transitions
$\alpha\beta$ and $\alpha\gamma$ at $q_1q'_1$, which are included in our PAA.
For the state $q_2q'_2$, it can be shown that the behaviours that are not
present in the PAA are in fact infeasible. There are two possible behaviors at
$q_2$: $i \geq len$ ($\alpha$) and $i<len$ ($\beta$); similarly, there are two
behaviors at $q'_2$:  $len'=0$ ($\gamma$) and $len'\neq0$ ($\delta$). Thus
there are four possible behaviors at $q_2q'_2$:  $\alpha \gamma$, $\alpha
\delta$, $\beta \gamma$, and $\beta \delta$. Since the behaviors $\alpha
\gamma$ and $\beta \delta$ are already included in the PAA, showing that
$\alpha \delta$ and $\beta \gamma$ are infeasible is sufficient. Observe that
at state $q_2q'_2$, the predicate $len-i=len' \land i \leq len$ is an
invariant. Since $i \geq len \land len'\neq0 \land len-i=len' \land i \leq len$ is
unsatisfiable, the behavior $\alpha \delta$ is infeasible. Similarly, $i<len
\land len'=0 \land len-i=len' \land i \leq len$ is unsatisfiable which implies
$\beta\gamma$ is infeasible.

We now justify why $len-i=len' \land i \leq len$ is an invariant at $q_2q'_2$.
Initially at $q_1q'_1$, $len$ and $len'$ are same and non-negative. There are
two ways to reach $q_2q'_2$ from $q_1q'_1$ depending on the parity of $len$. If
$len$ is even, both $len$ and $len'$ remain intact and $i$ is initialized to
$0$. If $len$ is odd, there is no change in $len$ and $i$ becomes $1$, however,
$len'$ is decreased by $1$. Therefore, $len-i=len' \land i \leq len$ is
initially true at $q_2q'_2$. Now, we prove the consecution. Assume at any step,
the predicate $len-i=len' \land i \leq len$ holds. Since the self-loop at
$q_2q'_2$ executes $b$ twice and $c'$ once, it preserves the satisfiability of
$len-i=len' \land i \leq len$: $i$ increases by $2$ and $len'$ decreases by
$2$. Therefore, it's an invariant at $q_2q'_2$. Note that $\omega = \omega'$
holds at $q_2q'_2$, which is further propagated to $q_3q'_3$ via transition
$c;d'$. It concludes that the two programs are equivalent since the content of
the arrays or final heaps are same.

\subsection{Soundness of our approach}
\label{subsec:soundness}

Our approach is sound by construction. An edge is added in the PAA if and only
if its source and target states are indeed connected through the
transition-label.  The choice of alignment predicates, and the inherent
incompleteness of the technique, may sometimes result in a PAA that's
insufficient to establish equivalent (for example, if it does not capture all
possible program behaviors). However, if a PAA and the learned invariants
logically establish the equivalence, the programs are indeed equivalent.

\section{Illustration on another example: arrayInsert}
\label{sec:arrayinsert}


We underline the usefulness of our direct construction, as compared to the
trace-based technique, using another example borrowed from~\cite{pdsc-cav19}.
Consider two copies, $f$ and $g$, of a program \texttt{arrayInsert}, as shown
in Fig.~\ref{fig:appendix-arrayinsert-code}; their CFGs are shown in
Fig.~\ref{fig:ap-arrayinsert-cfg}. The program takes 3 input parameters:
array $A$, its length \emph{len} and an integer $h$. The precondition under
which the equivalence is to be established is $A=A'$ and $len=len'$. The
variables $h$ and $h'$ are unconstrained by the precondition.

\begin{figure}[h]
\subcaptionbox{Copy $f$\label{subfig:appendix-code-f}}[%
    0.49\linewidth 
]%
{%
\begin{lstlisting}[label={lst:appendix-f-code}, caption={}, style=customc]
    arrayInsert(A, len, h)
    {
      i = 0;
      while (i<len && A[i]<h) 
        i++;
	
      len = shift_array(A,i,1);
      A[i] = h;

      while (i<len) { i++; }
      return i;
    }
\end{lstlisting}%
}%
\noindent
\subcaptionbox{Copy $g$\label{subfig:appendix-code-g}}[%
    0.40\linewidth 
]%
{%
\begin{lstlisting}[label={lst:appendix-g-code}, caption={}, style=customc]
arrayInsert(A, len, h)
{
   i = 0;
   while (i<len && A[i]<h) 
       i++;
	
   len = shift_array(A,i,1);
   A[i] = h;

   while (i<len) { i++; }
   return i;
}
\end{lstlisting}%
}%
\caption[]{Copies of \texttt{arrayInsert} program}
\label{fig:appendix-arrayinsert-code}
\end{figure}

The task here is to insert $h$ at its appropriate position in the sorted array
$A$, with the underlying assumption that $h$ is sensitive information and the
place where it is inserted must not be leaked. To achieve this, the programs
have a proxy loop towards the end, to move the counter $i$ to the end,
independent of the position where $h$ was inserted. The postcondition for
equivalence is that the output $i$ is the same for both the programs.

\begin{wrapfigure}{l}{0.65\textwidth}
    \centering
\begin{subfigure}[t]{0.27\textwidth}
\includegraphics[width=\linewidth]{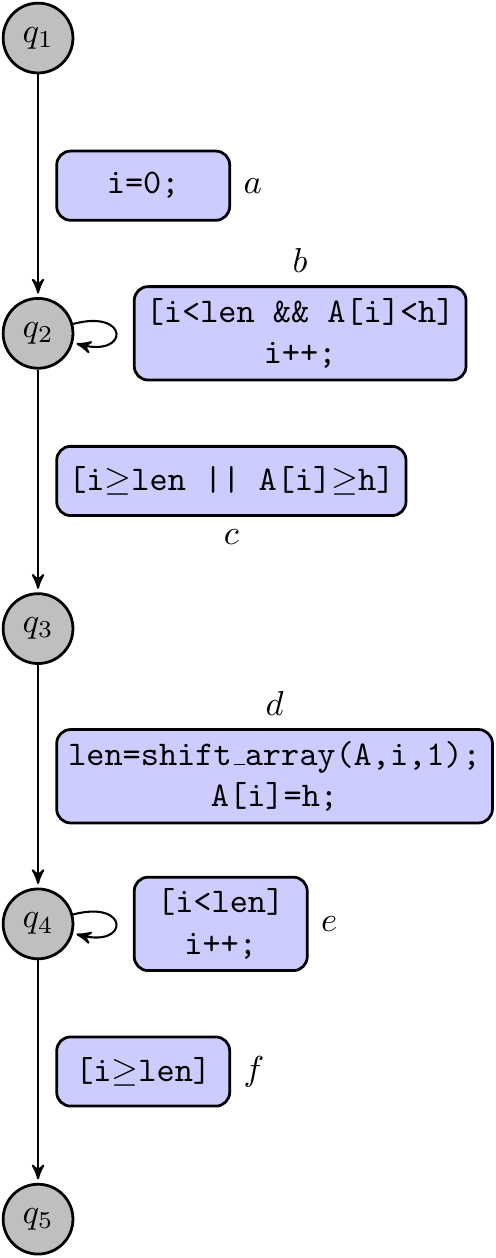}
 \caption{Program $f$ automaton\label{subfig:ap-arrayinsert-cfg-f}}		
 	\end{subfigure}
 \quad
 \hspace{1pt}
\begin{subfigure}[t]{0.28\textwidth}
\includegraphics[width=\linewidth]{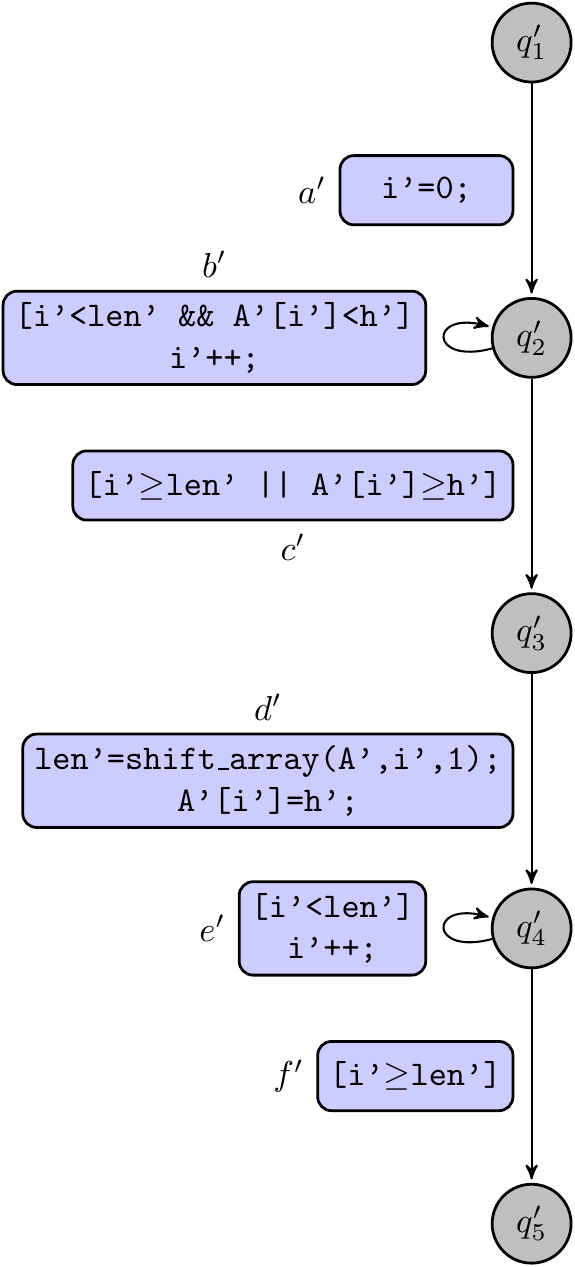}
 \caption{Program $g$ automaton\label{subfig:ap-arrayinsert-cfg-g}}		
 	\end{subfigure}
\caption{Control flow graphs for  Fig.~\ref{fig:appendix-arrayinsert-code} programs}
    \label{fig:ap-arrayinsert-cfg}
\end{wrapfigure}

\begin{figure}
\centering
\subcaptionbox{Trace based PAA for $h=h'$ or $h$ an $h'$ are inserted at  same positions\label{subfig:ap-trace-paa-1}}{\includegraphics[width=0.75\linewidth]{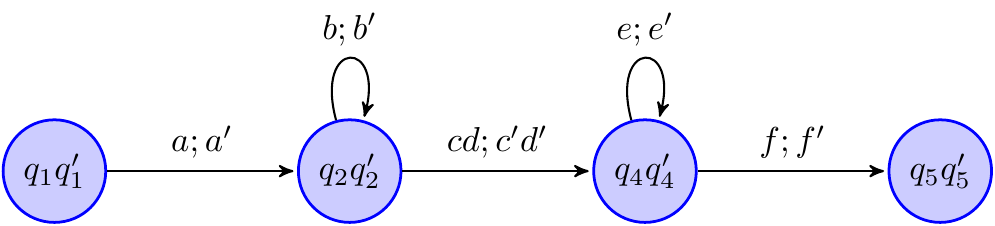}}
\subcaptionbox{Trace based PAA for $h>h'$ where $h$ and $h'$ are inserted  at different positions\label{subfig:ap-trace-paa-2}}{\includegraphics[width=0.8\linewidth]{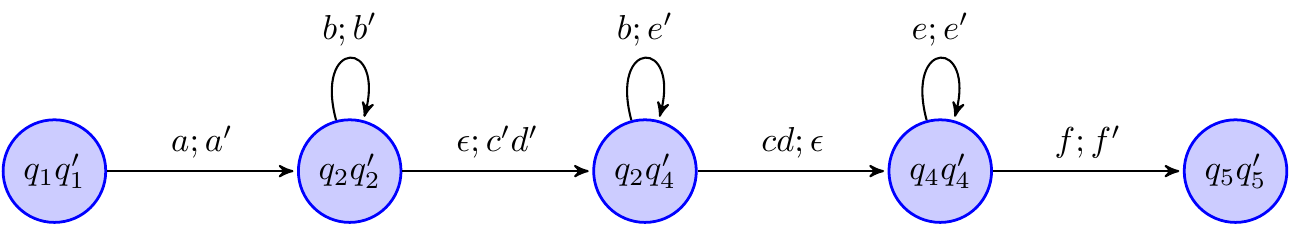}}
\subcaptionbox{Trace based PAA for $h<h'$ where $h$ and $h'$ are inserted at different positions\label{subfig:ap-trace-paa-3}}{\includegraphics[width=0.8\linewidth]{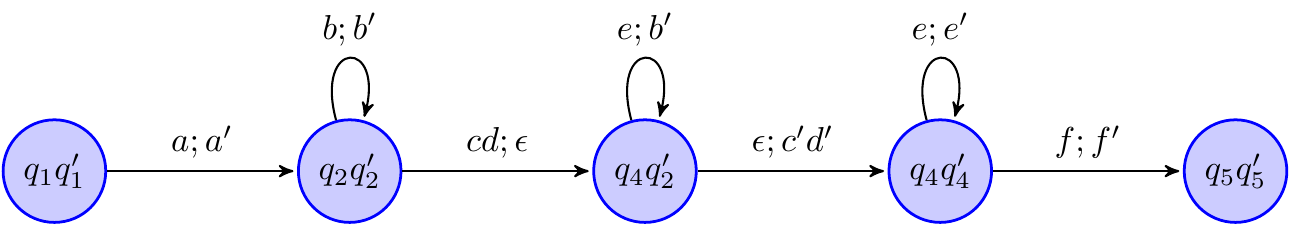}}
\caption{Trace PAAs for programs in Fig.~\ref{fig:ap-arrayinsert-cfg}}
\label{fig:ap-trace-paas}
\end{figure}

\begin{figure}
\centering
\includegraphics[width=0.6\linewidth]{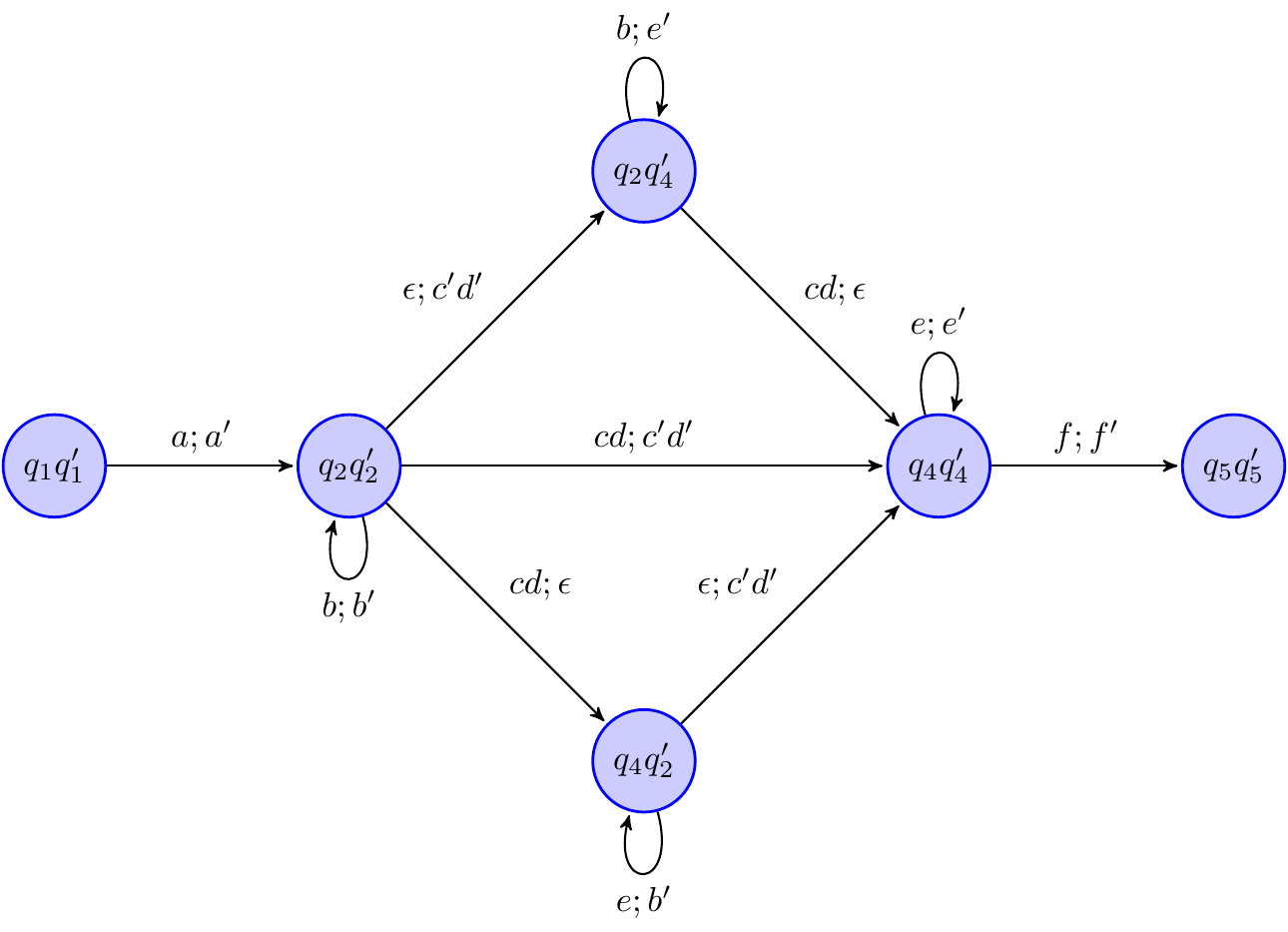}
\caption{Directly constructed PAA for programs in Fig.~\ref{fig:ap-arrayinsert-cfg}}
\label{fig:ap-our-paa}
	\end{figure}

Naturally, in this case, the predicate $i=i'$ appears to be a good candidate
for the alignment predicate $\mathcal{P}_{align}$ to construct a PAA.  There
are 3 scenarios based on the values of parameter \texttt{h} across both copies:
(i) $h=h'$ or both inserted at the same position in respective arrays, (ii)
$h<h'$ where $h$ and $h'$ are inserted at different positions, and (iii) $h>h'$
and both are inserted at different positions. The trace-based technique in
~\cite{spa-pldi19} would require a different pair of executions for computing
the trace alignment in each scenario. Fig.~\ref{fig:ap-trace-paas}
illustrates the program alignment automata constructed for each of these cases.
Absence of any of the pairs would lead to missing behaviors in the final PAA.
In contrast, our approach gives the PAA shown in Fig.~\ref{fig:ap-our-paa}.
We argue that this PAA observes each scenario and overapproximates all
behaviors.

Consider the initial state $q_1q'_1$: each of $q_1$ and $q'_1$ has one outgoing
transition with its guard predicate as $true$ (say, $\alpha$ and $\alpha'$
resp.). Hence there is only one transition $\alpha\alpha'$ at $q_1q'_1$, which
is included in the PAA. Now, let us consider the state $q_2q'_2$. We show that
the behaviours that are not present at $q_2q'_2$ are actually infeasible.  The
same argument can be extended to rest of the states in similar manner. There
are two behaviors possible at $q_2$: ($i<len \land A[i]<h$) (say, $\alpha$),
($i\geq len \lor A[i]\geq h$) ($\lnot \alpha$).  Similarly, $q'_2$ has two
possible behaviors: ($i'<len' \land A'[i']<h'$) (say, $\gamma$), ($i'\geq len'
\lor A'[i']\geq h'$) ($\lnot\gamma$). This leads to a total of four possible
behaviors at $q_2q'_2$: $\alpha\gamma$, $\lnot\alpha\gamma$,
$\alpha\lnot\gamma$, and $\lnot\alpha\lnot\gamma$, as shown below. The
alignment predicate $\mathcal{P}_{align}$ is $i=i'$, and $len=len'$ is a loop
invariant at $q_2q'_2$.

%

\begin{enumerate}
	\item $\alpha\gamma$: ($i<len \land A[i]<h$) $\land$ ($i'<len' \land A'[i']<h'$)
	\vspace{7pt}
	\item $\alpha\lnot\gamma$: ($i<len \land A[i]<h$) $\land$ ($i'\geq len' \lor A'[i'] \geq h'$)
	\begin{enumerate}
	    \item ($i<len \land A[i]<h$) $\land$ $i'\geq len'$ 
	    \item ($i<len \land A[i]<h$) $\land$ $A'[i'] \geq h'$
	\vspace{7pt}
	\end{enumerate}
	\item $\lnot\alpha\gamma$: ($i \geq len \lor A[i] \geq h$) $\land$ ($i'<len' \land A'[i']<h'$)
	\vspace{7pt}
	\begin{enumerate}
		\item $i \geq len$ $\land$ ($i'<len' \land A'[i']<h'$)
		\item $A[i] \geq h$ $\land$ ($i'<len' \land A'[i']<h'$)
	\vspace{7pt}
	\end{enumerate}
	\item $\lnot\alpha\lnot\gamma$: ($i \geq len \lor A[i] \geq h$) $\land$ ($i'\geq len' \lor A'[i'] \geq h'$)
	\begin{enumerate}
	\item $i \geq len$ $\land$ $i'\geq len'$
	\item $i \geq len$ $\land$  $A'[i'] \geq h'$
	\item $A[i] \geq h$ $\land$ $i'\geq len'$
	\item $A[i] \geq h$ $\land$ $A'[i'] \geq h'$
	\end{enumerate}
\end{enumerate}

\textbf{Case 1} corresponds to the self-loop $b;b'$ at $q_2q'_2$ which is
included in the PAA. 

\textbf{Case 2a} shows the predicate $i < len~\land~i' \geq len'$, which is
not satisfiable. The reason is that the alignment predicate $i=i'$ holds at
$q_2q'_2$ and $len=len'$ is a loop invariant. This infeasible transition is not
present in the PAA, which satisfies our requirement.

\textbf{Case 2b} represents the transition $q_2q'_2 \xrightarrow{\epsilon;c'd'}
q_2q'_4$ in our PAA. It is noteworthy that this transition is not included in
the automaton from the trace-based construction
(Figures~\ref{subfig:ap-trace-paa-1} and~\ref{subfig:ap-trace-paa-3}).

\textbf{Case 3a} is not a part of our PAA as well. The predicate $i \geq
len~\land~i' < len'$ is unsatisfiable, therefore, the transition is infeasible.

\textbf{Case 3b} is associated with the transition $q_2q'_2
\xrightarrow{cd;\epsilon} q_4q'_2$ in our PAA. However, this transition is not
included in the trace-based automata in Figures~\ref{subfig:ap-trace-paa-1}
and~\ref{subfig:ap-trace-paa-2}.

\textbf{Cases 4a, 4b, 4c, 4d} correspond to the transition $q_2q'_2
\xrightarrow{cd;c'd'} q_4q'_4$, which is a part of our program alignment
automaton.

It can similarly be argued that the program alignment automaton has all
possible behaviours at every state. Further, notice that $i=i'$ is an alignment
predicate, which holds at each state of the PAA by construction. In particular,
it holds at the exit state ($q_5q'_5$), and thus the PAA establishes
equivalence of the copies $f$ and $g$.

\section{Multiple Alignment Predicates and Disjunctive Invariants}
\label{sec:features}

Intuitively, a PAA is \emph{good} (in other words, \emph{useful} in making the
equivalence proof easier) if it can make the programs align at multiple
locations, i.e. if there are many intermediate nodes. In the worst case, if the
programs align only in the beginning, then the PAA cannot make the proof any
easier (than self-composing the programs and checking).

Consider two PAAs $A$ and $A'$ for alignment predicates $\mathcal{P}_{align}$
and $\mathcal{P'}_{align}$ respectively. If the number of reachable nodes in
$A$ is more than in $A'$, then $A$ is considered better aligned, which
certainly depends on the chosen alignment predicate. As an optimization, we can
parallelize computing transitions for multiple predicates and maintain multiple
transition sets.  Additionally, we can discard computing transitions for the
predicates that have significantly less number of reachable nodes than the
other. Multiple alignment predicates can also help in suggesting disjunctive
invariants.  For example, consider functions $f$ and $g$ shown in
Figures~\ref{subfig:disjunct-code-f} and~\ref{subfig:disjunct-code-g}. They
take two input parameters, $h$ and $cons$, and define two local variables
$y$, $z$. The function has a branching based on the value of $h$ -- the first
branch corresponds to the case $h > 100$ while the other is taken when $h \leq
100$. Now, assume two alignment predicates: $p_1 \deq z = z' + cons$ and
$p_2 \deq y = y' + cons$.  Recall that the alignment predicate is, by
assumption, true at initial state.  It is easy to observe that $p_1$ helps in
aligning first branch ($h > 100$) whereas $p_2$ assists in the alignment of the
other branch ($h \leq 100$). The predicate $p_1 \wedge p_2$ fails to align
either of the branches, whereas $p_1 \vee p_2$ helps in aligning both the
branches.

\begin{figure}[!ht]
\subcaptionbox{Function f\label{subfig:disjunct-code-f}}[%
    0.45\linewidth 
]%
{%
\begin{lstlisting}[label={lst:disjunct-f-code}, caption={}, style=customc]
    int f(int h, int cons)
    {
      int y = 2*h + cons;
      int z = 2*h + cons;

      if (h > 100) {y = 0;}
      else {z = 0;}

      while (h !=0) {
        if (y == 0) {z--;}
        else {y--;}
        h--;
      }
  
      if (y == 0) {return z;}
      return y;
    }
\end{lstlisting}%
}%
\noindent
\subcaptionbox{Function g\label{subfig:disjunct-code-g}}[%
    0.4\linewidth 
]%
{%
\begin{lstlisting}[label={lst:disjunct-g-code}, caption={}, style=customc]
    int g(int h, int cons)
    {
      int y = 2*h;
      int z = 2*h;

      if (h > 100) {y = 0;}
      else {z = 0;}

      while (h !=0) {
        if (y == 0) {z--;}
        else {y--;}
        h--;
      }

      if (y == 0) {return (z + cons);}
      return (y + cons);
    }
\end{lstlisting}%
}%
\caption[]{Multiple alignment predicates and their disjunction}
\label{fig:disjunct-code-f-g}
\end{figure}

\section{Related Work}
\label{sec:related}

Our work is closely related to and inspired by~\cite{spa-pldi19}, in that we
also use an alignment predicate to construct a program alignment automata that
semantically aligns the programs for equivalence check. However, our technique
constructs the PAA directly, without needing test cases or execution traces.
Our construction is similar in spirit to~\cite{DBLP:conf/aplas/DahiyaB17},
which builds a product program without using test cases, but it requires the
branching condition of one program to match that of the other. It also fails to
explore many-to-many relationship among paths of component programs, which we
do by constructing regular expressions and looking for suitable instantiations
of them. Another technique, CoVaC~\cite{CoVaC}, geared towards translation
validation, constructs a cross-product of two programs to ensure that
optimizing compiler transformations preserve program semantics. However, it
restricts the domain of transformations such that the optimized program is
consonant (structurally similar) to the source program.

Data-driven equivalence checking~\cite{DBLP:conf/oopsla/0001SCA13} tries to
find an inductive proof of loop equivalence in the domain of compiler
optimizations by inferring \emph{simulation relations} based on execution
traces and equality checking of the machine states. Since its goal is to align
loops, the technique is not suitable for the example in
Fig.~\ref{fig:code-cfg-f-g}. Other related techniques include those that prove
equivalence of loop-free
programs~\cite{1613165,10.1145/1086228.1086284,Feng2002AutomaticFV,10.1145/337292.337339,10.1007/11513988_20},
or programs with finite unwindings of loops or finite input
domains~\cite{10.1007/978-3-642-22110-1_55,10.1145/1453101.1453131,10.1007/978-3-642-31424-7_54,Jackson1994SemanticDA}.
There are also techniques that require some knowledge of the transformations
performed~\cite{eqsat-opt,10.1145/945885.945888} or the order of
optimizations~\cite{10.1007/BFb0054170,10.1145/349299.349314,GOLDBERG200553}.
In contrast, our approach can work with loops as well as in a black-box setting
where knowledge about the syntactic difference in the programs is not
available.

\section{Conclusion and Future Work}
\label{sec:conc}

We presented an algorithm for building program alignment automata, addressing
the equivalence checking problem for two programs. Our algorithm works directly
on the automaton of the individual programs, without needing any test cases or
making any unrealistic assumptions. Developing a prototype tool that implements
this algorithm is an immediate future work. In particular, it would be useful
to explore heuristics that make the technique scale in practice. For example,
by eagerly discarding states, transitions, and alignment predicates that are
not leading to a good alignment automaton. An aggressive reduction of the
product states may also help gain efficiency, though it may come at the cost of
completeness (i.e. the PAA missing some feasible behaviors).

\bibliographystyle{abbrv}
\bibliography{paper}

\end{document}